\documentclass[aps,prb,twocolumn,superscriptaddress,showpacs]{revtex4}

\usepackage{graphicx}
\usepackage{dcolumn}
\usepackage{bm}
\usepackage{amsfonts}
\usepackage{amsmath}

\begin{document}

\title{Deviation from the Wiedemann-Franz law induced by nonmagnetic 
impurities in overdoped La$_{2-x}$Sr$_x$CuO$_4$}

\author{X. F. Sun}
\email{xfsun@ustc.edu.cn}

\affiliation{Hefei National Laboratory for Physical Sciences at 
Microscale, University of Science and Technology of China, Hefei, 
Anhui 230026, P. R. China} 

\affiliation{Central Research Institute of Electric Power 
Industry, Komae, Tokyo 201-8511, Japan}

\author{B. Lin}
\affiliation{Hefei National Laboratory for Physical Sciences at 
Microscale, University of Science and Technology of China, Hefei, 
Anhui 230026, P. R. China}

\author{X. Zhao}
\affiliation{Department of Astronomy and Applied Physics, 
University of Science and Technology of China, Hefei, Anhui 
230026, P. R. China}

\author{L. Li}
\affiliation{Hefei National Laboratory for Physical Sciences at
Microscale, University of Science and Technology of China, Hefei,
Anhui 230026, P. R. China}

\author{Seiki Komiya}
\affiliation{Central Research Institute of Electric Power
Industry, Komae, Tokyo 201-8511, Japan}

\author{I. Tsukada}
\affiliation{Central Research Institute of Electric Power
Industry, Komae, Tokyo 201-8511, Japan}

\author{Yoichi Ando}
\affiliation{Institute of Scientific and Industrial Research,
Osaka University, Ibaraki, Osaka 567-0047, Japan}

\date{\today}

\begin{abstract}

To investigate the validity of the Wiedemann-Franz (WF) law in 
disordered but metallic cuprates, the low-temperature charge and 
heat transport properties are carefully studied for a series of 
impurity-substituted and carrier-overdoped 
La$_{1.8}$Sr$_{0.2}$Cu$_{1-z}$$M$$_z$O$_4$ ($M$ = Zn or Mg) 
single crystals. With moderate impurity substitution 
concentrations of $z$ = 0.049 and 0.082 ($M$ = Zn), the 
resistivity shows a clear metallic behavior at low temperature 
and the WF law is confirmed to be valid. With increasing impurity 
concentration to $z$ = 0.13 ($M$ = Zn) or 0.15 ($M$ = Mg), the 
resistivity shows a low-$T$ upturn but its temperature dependence 
indicates a finite conductivity in the $T \rightarrow 0$ limit. 
In this weakly-localized metallic state that is intentionally 
achieved in the overdoped regime, a {\it negative} departure from 
the WF law is found, which is opposite to the theoretical 
expectation. 

\end{abstract}

\pacs{74.25.Fy, 74.72.Dn}

\maketitle

\section{Introduction}

The Wiedemann-Franz (WF) law, which relates the thermal 
conductivity $\kappa$ and the electrical conductivity $\sigma$ 
through a simple formula,
\begin{equation}
\kappa/\sigma T = L, \label{WF}
\end{equation}
where $L$ is called the Lorentz number and is given by the 
Sommerfeld's value $L_0 = 2.44 \times 10^{-8}$ W$\Omega$/K$^2$, 
is a robust signature of a Fermi liquid. It relies on the 
single-particle description of the transport properties and a 
purely elastic electron scattering. Although the WF law is 
usually not obeyed at $T \neq$ 0 because of the importance of 
inelastic scattering, it must be obeyed in the Fermi liquid at $T 
=$ 0 where the electrons are elastically scattered by static 
disorders. The examination of the WF law at $T \rightarrow 0$ by 
using sub-Kelvin charge and heat transport measurements can 
therefore provide a direct judgement on the nature of the normal 
state of high-$T_c$ cuprates, which is a prime example of the 
strongly-correlated electron systems challenging the description 
of Fermi liquids. It has been found in the overdoped 
Tl$_2$Ba$_2$CuO$_{6+\delta}$ (Tl2201) that the WF law is 
perfectly obeyed, which indicated a Fermi-liquid ground state of 
the overdoped cuprates.\cite{Proust1} 

It was recently predicted that the WF law is violated in 
disordered interacting electron systems,\cite{Raimondi, Niven, 
Catelani} in which the Coulomb interactions are likely to bring 
some additional scattering and impede the charge transport more 
efficiently. A recent study of the low-$T$ transport of 
Bi$_{2+x}$Sr$_{2-x}$CuO$_{6+\delta}$ (Bi2201),\cite{Proust2} in 
which a positive departure from the WF law ($\kappa/\sigma T > 
L_0$) appeared at optimum doping and became more pronounced in 
the underdoped regime, was argued to support the theoretical 
prediction because the disorder in Bi2201 increases with 
decreasing doping level due to its peculiar chemistry. However, 
in previous studies, both on the overdoped Tl2201 and on the 
underdoped Bi2201, the normal state was achieved by applying very 
strong magnetic field, which could induce some kind of SDW/CDW 
order.\cite{Hoffman, Lake, Khaykovich} Such a competing order may 
cause some uncertainty regarding whether the properties of the 
true ground state is indeed detected.\cite{Kivelson, Sonier} It 
would therefore be useful if one could study the validity of the 
WF law in cuprates whose normal states is obtained in a way other 
than by applying a strong magnetic field. 

Introducing impurities or defects into CuO$_2$ planes is a viable 
way of suppressing the superconductivity and obtaining the normal 
state. Because of the non-$s$-wave symmetry of cuprates, 
nonmagnetic impurities like Zn or Mg are found to destroy the 
superconductivity quite strongly, particularly for the underdoped 
cuprates.\cite{Fukuzumi, Komiya1, Hanaki, Raffo} Furthermore, the 
impurity substitution was found to show different impacts on the 
physical properties in the underdoped and the overdoped cuprates, 
signifying a remarkable difference in the corresponding ground 
states.\cite{Fukuzumi} 

It was found that in underdoped La$_{2-x}$Sr$_x$CuO$_4$ 
(LSCO) and Bi$_2$Sr$_{2-x}$La$_x$CuO$_{6+\delta}$ (BSLCO), a 
low-$T$ resistivity divergence shows up immediately after the 
superconductivity is completely suppressed by a slight Zn 
substitution ($\sim$ 2\%).\cite{Fukuzumi, Komiya1, Hanaki} The 
resistivity divergence is even stronger than the log(1/$T$) 
function, which is a well-known low-$T$ behavior of 
high-magnetic-field-induced insulating state,\cite{Ando, Ono} and 
therefore indicates an insulating ground state.\cite{Komiya1, 
Hanaki} The magnetoresistance data suggested that the Zn-induced 
insulating state is likely due to Kondo effect associated with 
impurity-induced magnetism.\cite{Hanaki} In any event, the 
normal state of these disordered and underdoped cuprates is 
apparently insulating and it is actually not meaningful to test 
the validity of the WF law for these samples. 

In optimally-doped or overdoped regimes, on the other hand, a 
slight Zn substitution ($\sim$ 4\%) was found to induce a simply 
metallic state when the superconductivity is 
suppressed,\cite{Fukuzumi} which is believed to demonstrate the 
Fermi liquid ground state of the overdoped cuprates. However, it 
was also found that the heavy electron irradiation (to introduce 
point defects and to destroy the superconductivity similarly to 
Zn substitution) in the overdoped Tl2201 can lead to a low-$T$ 
resistivity upturn,\cite{Albenque1} so the occurrence of electron 
localization in the overdoped regime appears to depend on the 
level of disorder. The temperature dependence of the conductivity 
suppression $\Delta \sigma$ (compared to the extrapolated 
residual conductivity from the high-$T$ data) in the 
heavily-electron-irradiated Tl2201 had a ln$T$ variation and was 
in rough quantitative agreement with the two-dimensional (2D) 
weak localization theory. Once the Anderson transition into the 
insulating state takes place and the resistivity shows an 
exponentially divergent behavior due to strong disorder, even in 
the overdoped regime the WF law is meaningless. However, in the 
moderately disordered regime in which the superconductivity is 
suppressed by disorder but the resistivity is not really 
divergent, it becomes possible to examine the peculiar electron 
transport properties through the WF law without introducing the 
possible complications from the magnetic-field-induced effects. 

In this work, by substituting Zn or Mg for Cu in 
La$_{1.8}$Sr$_{0.2}$Cu$_{1-z}$$M$$_z$O$_4$ ($M$ = Zn or Mg) 
(LSCMO), we are able to obtain a series of non-superconducting 
overdoped LSCO. When the superconductivity is just suppressed at 
Zn concentration of $z$ = 0.049 and 0.082, the ground state is 
clearly metallic and the WF law is found to hold reasonably well, 
pointing to the Fermi-liquid state of the overdoped cuprates. 
With higher concentration of Zn ($z$ = 0.13) or Mg ($z$ = 0.15), 
a low-$T$ resistivity upturn shows up. It is found that the 
$T$-dependence of the conductivity obeys a $T^{1/2}$ law and 
therefore indicates a metallic ground state at $T \rightarrow 0$. 
This result, accompanied with the negative magnetoresistance, 
suggests a three-dimensional (3D) weak localization behavior. 
However, the low-$T$ heat transport data indicate a surprisingly 
{\it negative} deviation from the WF law ($\kappa_0/\sigma T < 
L_0$) that is opposite to the theoretical prediction for the 
disordered interacting electron systems. This new result 
demonstrates yet another unconventional feature of the mysterious 
normal state of the high-$T_c$ cuprates.

\section{Experiments}

High-quality La$_{1.8}$Sr$_{0.2}$Cu$_{1-z}$$M$$_z$O$_4$ ($M$ = Zn 
or Mg) single crystals are grown by a traveling-solvent 
floating-zone method with similar processes to that reported in 
Ref. \onlinecite{Komiya2}. The actual Zn or Mg concentration $z$ 
is determined by the inductively-coupled plasma atomic-emission 
spectroscopy (ICP-AES). All the samples are carefully checked by 
using the X-ray Laue photograph and cut precisely along the 
crystallographic axes. The in-plane resistivity and the thermal 
conductivity are measured actually along the $a$ axis. To remove 
the oxygen defects which tend to occur in overdoped LSCO, the 
crystals are annealed at 800$^{\circ}$C in sealed quartz tubes 
with 4 atm oxygen pressure for at least 5 days, followed by rapid 
quenching to the liquid nitrogen temperature. Several pieces of 
crystals used for transport measurements at each impurity 
concentration are labelled ``A", ``B", ``C", etc. The in-plane 
resistivity $\rho_{ab}$ is measured using a standard ac 
four-probe method in a $^4$He cryostat from 1.4 to 300 K and 
using a LR-700 ac-resistance bridge in a dilution refrigerator 
from 80 to 900 mK, respectively. In the latter case, excitation 
voltages are minimized to avoid observable self-heating effect. 
The thermal conductivity measurement in the sub-Kelvin region is 
done by a conventional steady-state ``one heater, two 
thermometer" technique with RuO$_2$ sensors in the dilution 
refrigerator.\cite{Sun1} 

\section{Results and Discussion}

\subsection{Low-temperature resistivity}

\begin{figure}
\includegraphics[clip,width=8.5cm]{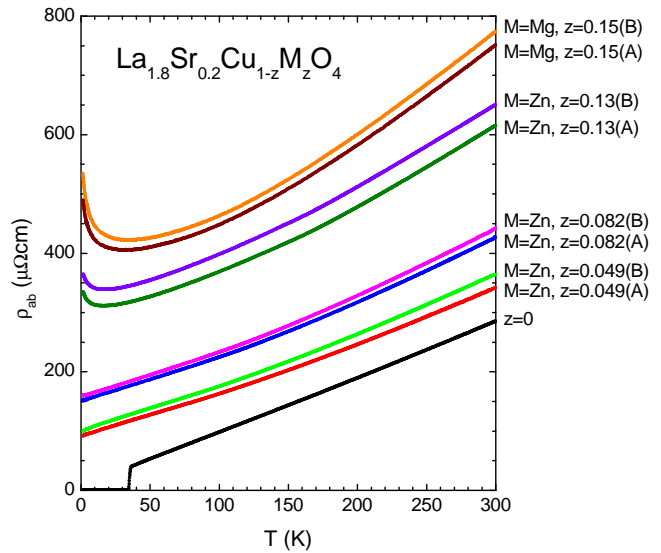}
\caption{(Color online) Temperature dependences of the in-plane 
resistivity of several Zn- and Mg-substituted LSCO single 
crystals. Two pieces of samples (labelled A and B) for each 
impurity concentration are measured for checking the 
uncertainties of the transport data. The small discrepancy ($<$ 
10\%) of the resistivity data between different samples at the 
same impurity concentration indicates a high quality of the 
crystals.}
\end{figure}

Figure 1 shows the in-plane resistivity $\rho_{ab}$ of several 
Zn- and Mg-substituted LSCO single crystals with Sr content $x$ = 
0.2. The pristine LSCO sample shows a good $T$-linear behavior of 
$\rho_{ab}$ below 300 K and a sharp superconducting transition at 
34.5 K. With increasing Zn concentration to $z$ = 0.049 and 
0.082, the linearity of the resistivity becomes a bit worse but 
obviously a metallic behavior (with positive temperature 
coefficient) persists down to very low temperature without 
displaying any superconducting transition. When Zn concentration 
increases to 0.13, the resistivity starts to show a weak upturn 
below $\sim$ 12 K. In Mg-substituted samples ($z$ = 0.15), the 
low-temperature upturn of the resistivity is pronounced. These 
data demonstrate the impact of impurity/disorder on the electron 
transport and superconductivity, all consistent with the 
literature.\cite{Alloul} In particular, the low-$T$ resistivity 
upturn in the heavily Zn- or Mg-substituted samples are 
qualitatively the same as that in the electron irradiated 
Tl2201.\cite{Albenque1} 

\begin{figure}
\includegraphics[clip,width=8.5cm]{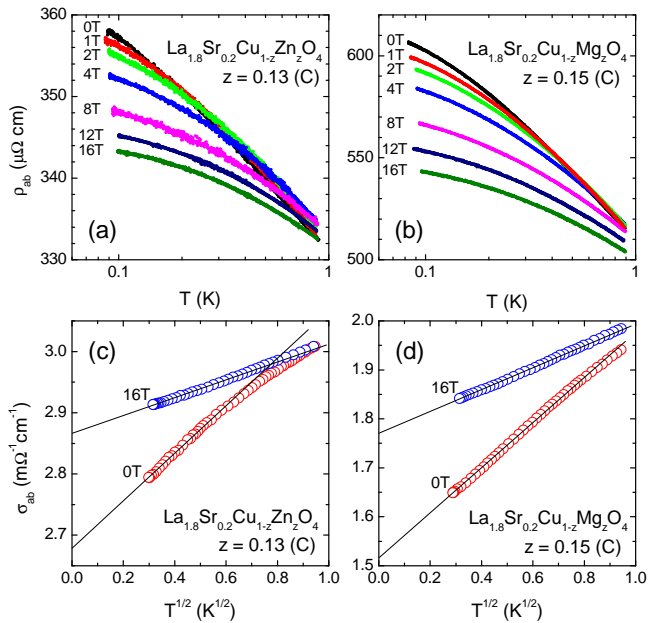}
\caption{(Color online) (a,b) Low-temperature resistivity of 
Zn-substituted ($z$  0.13) and Mg-substituted ($z$ = 0.15) LSCO 
single crystals in different magnetic fields (0--16 T and along 
the $c$ axis). (c,d) The electrical conductivity of the above two 
samples in 0 and 16 T fields, plotted in $\sigma_{ab}$ vs 
$T^{1/2}$. The solid lines are linear fittings to the 
low-temperature data.} 
\end{figure}

To further investigate the low-$T$ transport properties of highly 
Zn- and Mg-substituted LSCO, the in-plane resistivity is studied 
down to milli-Kelvin temperature and in magnetic fields from 0 to 
16 T ($\parallel c$). The detailed data are shown in Figs. 2(a) 
and 2(b) in log$T$ plots. First of all, the resistivity of these 
samples keeps increasing upon lowering temperature down to the 
lowest temperature ($\sim$ 80 mK). However, the increase in 
resistivity is apparently weaker than the log(1/$T$) function, 
which is a common insulating behavior in both the 
high-magnetic-field-induced normal state of underdoped high-$T_c$ 
cuprates and the lightly-doped non-superconducting 
YBa$_2$Cu$_3$O$_y$ (YBCO).\cite{Ando, Ono, Sun2} This observation 
suggests that the resistivity may not diverge in the 
zero-temperature limit in these samples. It is actually easy to 
find, as shown in Figs. 2(c) and 2(d), that the low-$T$ 
electrical conductivity of these samples follows a power law
\begin{equation}
\sigma_{ab} = \sigma_0 + \beta T^{1/2}, \label{sigma}
\end{equation}
where $\sigma_0$ and $\beta$ are temperature-independent 
coefficients. It follows from this temperature dependence that 
the resistivity $1/(\sigma_0 + \beta T^{1/2})$ remains finite at 
$T \to$ 0, which is a clear indication of a metallic ground state 
of these samples. Note that this $T$-dependence of resistivity is 
somehow different from the ln$T$ variation of $\Delta \sigma$ 
observed in the overdoped Tl2201 after significant electron 
irradiation.\cite{Albenque1} Since the measurements of Tl2201 
were not done at very low temperatures, a direct comparison 
between LSCMO and Tl2201 is not available now. Nevertheless, the 
2D weak localization picture proposed for Tl2201, which is based 
on the ln$T$ behavior of conductivity, may not be appropriate for 
LSCMO. Remember, the particular $T^{1/2}$ dependence of 
conductivity is known for the 3D weak localization.\cite{Mott, 
Lee} It is possible that the overdoped LSCO with strong disorder 
is effectively a 3D electron system, though the temperature 
dependence of the resistivity anisotropy suggests that clean LSCO 
is two dimensional for the whole superconducting compositions ($x 
<$ 0.30).\cite{Nakamura} 

Another important feature one may note in Fig. 2 is that the 
low-$T$ resistivity decreases in magnetic fields, that is, they 
show rather strong negative magnetoresistance. This immediately 
rules out the possible contamination of the superconducting 
fluctuations to the resistivity data and suggests that the weak 
negative temperature coefficient of resistivity is due to the 
intrinsic electron transport of the normal state. Furthermore, 
the negative magnetoresistance at low temperature also indicates 
that the weak localization is likely responsible for the 
resistivity upturn.\cite{Mott, Lee} 

As a summary of the electrical transport results, the resistivity 
upturn in the heavily Zn- and Mg-substituted and 
carrier-overdoped LSCO does not signify a metal-to-insulator 
crossover (MIC) but is attributed to an inelastic electron 
scattering by impurities/disorders in a metallic regime. 
Furthermore, the temperature dependence of the low-$T$ 
resistivity does not suggest a Kondo-like scattering mechanism, 
which is often playing a role in the charge transport in 
underdoped cuprates.\cite{Hanaki, Albenque2}

\subsection{Low-temperature thermal conductivity of simply metallic samples}

\begin{figure}
\includegraphics[clip,width=6.5cm]{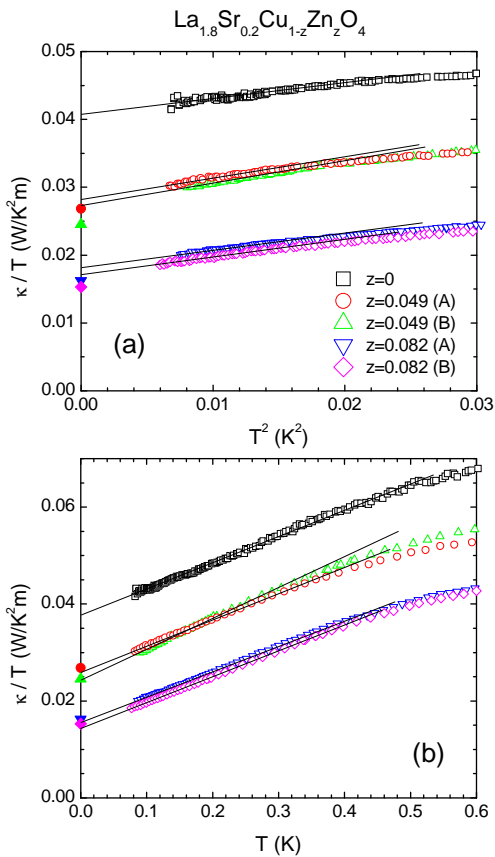}
\caption{(Color online) Low-temperature thermal conductivity of 
Zn-substituted LSCO single crystals, whose resistivity data are 
shown in Fig. 1. The data are plotted in (a) $\kappa/T$ vs $T^2$ 
and (b) $\kappa/T$ vs $T$. The solid lines are linear fittings to 
the low-temperature data in two different plots. Note the 
difference in the residual term obtained by the two kinds of 
analysis. The solid symbols at $T$ = 0 K indicate the values of 
$\kappa/T$ calculated by using the Wiedemann-Franz law with the 
residual resistivity obtained from Fig. 1. Note that the 
calculated $\kappa_0/T$ and the experimental $\kappa/T$ data are 
shown with the same shape of the symbols for the same samples.} 
\end{figure}

To further investigate the electron transport properties, the 
thermal conductivity $\kappa$ of Zn- and Mg-substituted LSCO 
samples shown in Fig. 1 have been measured down to 70 mK; Fig. 3 
shows the data for the superconducting 
La$_{1.8}$Sr$_{0.2}$CuO$_4$ sample and LSCMO samples with Zn 
concentration of 0.049 and 0.082, which showed the metallic 
resistivity behavior down to very low temperatures. Note that the 
low-$T$ thermal conductivity of all these samples can be well 
fitted either to 
\begin{equation}
\frac{\kappa}{T} = \frac{\kappa_0}{T} + bT^2, \label{T^2}
\end{equation}
or to
\begin{equation}
\frac{\kappa}{T} = \frac{\kappa_0}{T} + AT, \label{T}
\end{equation}
as shown in Figs. 3(a) and 3(b), respectively, but the 
temperature range for obtaining a good fit is much wider with 
using Eq. (\ref{T}). Looking at Fig. 3, one may notice that the 
data quality of these charge-carrier-overdoped LSCO samples are 
unprecedentedly high. In a previous study,\cite{Hawthorn1} some 
clear downturn of the low-$T$ thermal conductivity data was 
observed in the overdoped LSCO and was already discussed to be 
due to the electron-phonon decoupling,\cite{Smith1} whose adverse 
effects on probing the intrinsic electron thermal conductivity 
can be avoided if the contacts are good enough. 

Note that Eq. (\ref{T^2}) is commonly used for describing the 
low-$T$ thermal conductivity of superconducting cuprates, where 
the residual term $\kappa_0/T$ is the contribution of $d$-wave 
nodal quasiparticles and $bT^2$ is the phonon contribution in the 
boundary scattering limit.\cite{Taillefer, Takeya, Sun3, Sun4} In 
this regard, the present data of the pristine LSCO are in good 
harmony with the previous studies. For samples with Zn 
concentrations of 0.049 and 0.082, whose ground state is metallic 
as indicated by the in-plane resistivity, $\kappa_0/T$ is 
apparently coming from the electron heat transport as in ordinary 
metals. If the ground state is a Fermi liquid, one can expect 
that the values of the residual resistivity and $\kappa_0/T$ 
satisfy the WF law. From Fig. 3(a), it is clear that there is 
about 20\% difference between the $\kappa_0/T$ obtained from Eq. 
(\ref{T^2}) and the values calculated from the residual 
resistivity (see Fig. 1) using the WF law. Note that this 20\% 
difference is not mainly due to the experimental error, since the 
resistivity and thermal conductivity measurements are performed 
using the same samples and the same contacts. 

It was recently reported that Eq. (\ref{T}) is a good description 
for the low-$T$ thermal conductivity of Tl2201 and 
Bi$_2$Sr$_2$CaCu$_2$O$_{8+\delta}$.\cite{Hawthorn2, Sun1} As 
shown in Fig. 3(b), fitting the present thermal conductivity data 
to Eq. (\ref{T}) gives qualitatively the same behavior upon Zn 
substitution as that in Fig. 3(a), but the $\kappa_0/T$ from this 
fitting agree better (in fact, almost perfectly) with those from 
the WF law. This gives a circumstantial support to the validity 
of Eq. (4) in the overdoped regime. 

Compared to the underdoped LSCO, in which a modified formula with 
the phonon boundary scattering $\kappa/T = \kappa_0/T + 
b'T^{\alpha-1}$ ($\alpha$ = 2.5--3) was proposed to fit the data 
well,\cite{Hawthorn1} the second term $AT$ in Eq. (\ref{T}) is 
apparently not originating from the boundary scattering limit. 
One possible explanation for this unusual power is the effect of 
strong electron-phonon scattering on phonons. It has been known 
that a $T^2$ behavior of the phonon thermal conductivity can show 
up in both usual metals and $d$-wave 
superconductors.\cite{Berman, Smith2} In the case of a $d$-wave 
superconductor, the Fermi surface is anisotropicaly gapped 
except for the gap nodes, nearby where the quasiparticles are 
excited even at low temperature.\cite{ARPES, Hussey} Those 
so-called nodal quasiparticles dominate the transport 
properties.\cite{Hussey} For such a system, it was theoretically 
proposed that the $T^2$ behavior of the phonon thermal 
conductivity originates from the fact that the transverse phonons 
propagating along certain high-symmetry directions do not 
interact with the nodal quasiparticles.\cite{Smith2} However, the 
non-superconducting overdoped samples have a complete Fermi 
surface, and therefore a very different momentum distribution of 
electrons from that in $d$-wave superconductors is formed. 
Apparently, the electron-phonon scattering should differ 
significantly for the Zn-free and Zn-substituted samples, 
considering the distinct difference in the electronic structures 
between the superconducting and non-superconducting samples. 
Therefore, the $T^2$ coefficient is expected to be quite 
different in these two cases. Nevertheless, as shown in Fig. 3, 
the coefficient $A$ of the pristine (superconducting) and 
Zn-substituted (non-superconducting) LSCO are found to be nearly 
the same, which seems to rule out the possibility that the 
$T^2$-law comes from the electron-phonon scattering. In addition, 
one may notice that, when the electron-phonon scattering is 
strong, the electronic thermal conductivity $\kappa_e/T$ cannot 
be temperature independent and therefore Eq. (\ref{T}) is 
physically imprecise.\cite{Hawthorn2} 

Another possible origin of the $AT$ term is the phonon thermal 
conductivity with the dominant scattering of 
dislocations.\cite{Berman} This scenario is consistent with the 
observation that Eq. (\ref{T^2}) becomes a good description of 
the data at very low temperatures, because the dislocation 
scattering is likely to be smeared out at low enough 
temperatures. There is a possibility that some dislocation 
defects are produced in LSCO crystals when they are quenched to 
nitrogen temperature after high-temperature annealing.  

In any case, it appears that the WF law holds reasonably well 
(the deviation is no more than 20\%) in the overdoped LSCO when 
the normal state is obtained by a moderate impurity substitution. 
If one could take the validity of the WF law as a robust 
signature of a Fermi-liquid state, the present result apparently 
indicates a Fermi-liquid nature of the overdoped LSCO. This 
conclusion is essentially the same as that for the overdoped 
Tl2201, where the normal state was achieved by strong magnetic 
fields.\cite{Proust1} However, one should note that the 
resistivity of our moderately Zn-substituted samples actually 
shows a good $T$-linear behavior at low temperature, which has 
been taken as a characteristic non-Fermi-liquid 
feature.\cite{Cooper} Therefore, there is still some 
inconsistency between the interpretations of the WF law and the 
resistivity behavior in overdoped LSCO, which needs further 
clarification. 

\subsection{Low-temperature thermal conductivity for samples showing resistivity upturn}

\begin{figure}
\includegraphics[clip,width=8.5cm]{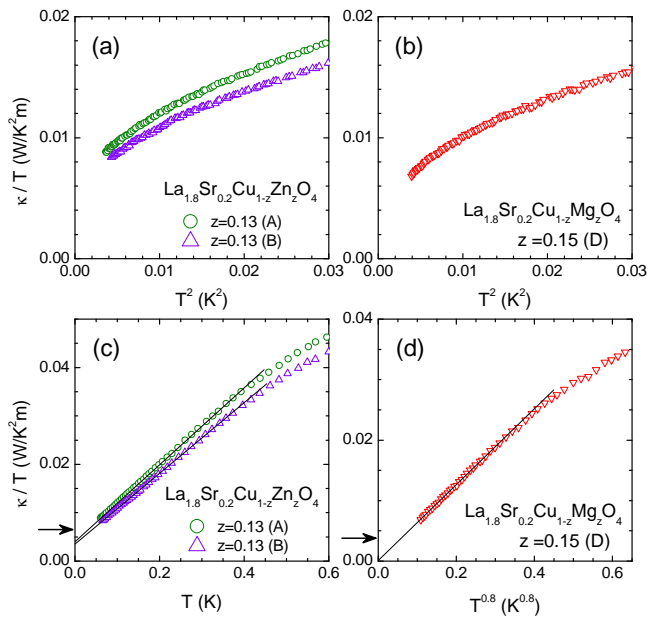}
\caption{(Color online) Low-temperature thermal conductivity of 
Zn-substituted ($z$ = 0.13) and Mg-substituted ($z$ = 0.15) LSCO 
single crystals. (a,b) Data are plotted in $\kappa/T$ vs $T^2$ 
and the linear relationship is not observed even in the lowest 
temperature regime. (c,d) Data are plotted in $\kappa/T$ vs $T$ 
and $\kappa/T$ vs $T^{0.8}$ for Zn- and Mg-substituted samples, 
respectively. Linear fittings to the low-temperature data are 
shown by solid lines. The arrows indicate the residual thermal 
conductivity $\kappa_0/T$ calculated  by using the 
Wiedemann-Franz law from the residual electrical conductivity 
$\sigma_0$ of the data shown in Fig. 2; note that they are very 
different from the extrapolated residual value indicated by the 
linear fitting to the experimental data.} 
\end{figure} 

The low-temperature thermal conductivity data of LSCMO with 
higher impurity concentration $z$ = 0.13 ($M$ = Zn) and 0.15 ($M$ 
= Mg) are shown in Fig. 4. From Figs. 4(a) and 4(b), it can be 
seen that the usual formula of Eq. (\ref{T^2}) is not able to fit 
the low-$T$ thermal conductivity data, that is, the experimental 
data plotted in $\kappa/T$ vs $T^2$ never show a linear behavior 
down to $\sim$ 60 mK; in fact, the curves always have a negative 
curvature. This phenomenon is very similar to what we observed in 
the lightly-doped non-superconducting YBCO,\cite{Sun2} in which 
case the heading of $\kappa/T$ to zero upon lowering temperatures 
and the log(1/$T$)-divergent resistivity demonstrated the 
insulating ground state. 

In the present case, we would like to point out at the beginning 
that the curvature of $\kappa/T$ vs $T^2$ plots is {\it not} 
likely due to the decoupling of electrons and 
phonons.\cite{Hawthorn1, Smith1} It is known that the necessity 
of making good contacts to obtain the intrinsic electronic 
thermal conductivity data becomes more severe when the electron 
term becomes larger. Comparing the data shown in Figs. 4 and 3, 
one can easily see that the pristine and lower Zn-substituted 
LSCO samples, which must have larger electron thermal 
conductivity than $z$ = 0.13 ($M$ = Zn) and 0.15 ($M$ = Mg) 
samples, actually do not show obvious sign of downturn curvature 
in their $\kappa/T$ vs $T^2$ plots.

Upon further analyzing the data, one can find that Eq. (\ref{T}) 
is still able to fit the thermal conductivity data of $z$ = 0.13 
($M$ = Zn) samples very well from 60 to 400 mK, as shown in Fig. 
4(c). Because of the even stronger curvature in the $\kappa/T$ 
data of $z$ = 0.15 ($M$ = Mg) sample, neither Eq. (\ref{T^2}) nor 
Eq. (\ref{T}) can fit the data; instead, the low-$T$ data of 
Mg-substituted LSCO can be described by a formula
\begin{equation}
\frac{\kappa}{T} = \frac{\kappa_0}{T} + A'T^{0.8}, \label{T^{0.8}}
\end{equation}
with the fitting parameter $\kappa_0/T$ = 0. 

As we have seen in Sec. III. A., the charge transport data 
indicate that in the heavily Zn- or Mg-substituted LSCO some 
inelastic scattering is taking place at low temperatures, causing 
a weak localization behavior. One direct consequence of this fact 
is that $\kappa_e/T$ is no longer independent of temperature in 
these samples, and the resistivity upturn naturally corresponds 
to a gradual decrease of $\kappa_e/T$ with lowering temperatures. 
Obviously, this invalidates Eq. (\ref{T}). Nevertheless, one can 
apparently fit the data for $z$ = 0.13 ($M$ = Zn) samples to Eq. 
(\ref{T}), which is probably due to the fact that the low-$T$ 
resistivity upturn ($<$ 10\% increase in $\rho_{ab}$ from 900 to 
80 mK) or the inelastic scattering are weak enough to make the 
deviations of $\kappa_e/T$ from a constant effectively 
negligible. In contrast, the $z$ = 0.15 ($M$ = Mg) samples show 
even stronger resistivity upturn and more significant inelastic 
scattering. It is likely that a considerable temperature 
dependence of the electronic term $\kappa_e/T$ is the cause of 
Eq. (\ref{T^{0.8}}), which is an approximate description of the 
low-$T$ thermal conductivity of Mg-substituted sample.  

Based on the linear fittings (shown in Fig. 4) to obtain the 
residual term $\kappa_0/T$, one can easily find an important 
result, that is, the WF law in these metallic samples are 
strongly violated. As shown in Figs. 4(c) and 4(d), the 
$\kappa_0/T$ values obtained from the extrapolation of the 
thermal conductivity data to $T$ = 0 K are much smaller than 
those predicted by the WF law using the residual conductivity 
shown in Figs. 2(c) and 2(d). Especially, the Mg-substituted LSCO 
crystal gives a $\kappa_0/T$ value that is essentially zero, 
indicating a ``thermal insulator" ground state, while the 
resistivity data suggest a finite residual term and a metallic 
ground state. It is surprising that the departure from the WF law 
is {\it negative}, which has never been observed in cuprates; for 
example, upon approaching MIC in the underdoped region of 
cuprates, the departure from the WF law was found to be positive. 
\cite{Proust2} Intriguingly, there is another metallic system 
which exhibits a negative deviation from the WF law: It is the 
heavy-fermion system CeCoIn$_5$ when tuned to its quantum 
critical point (QCP).\cite{Tanatar} In that materials, the 
violation of the WF law is accompanied by a $T$-linear 
resistivity and has been discussed to be due to an anisotropic 
destruction of the Fermi surface. Such a QCP-related origin of 
the WF-law violation is unlikely for LSCO, given that the 
violation is {\it not} observed in less disordered ($z$ = 0.049 
and 0.082) samples. In passing, it is useful to note that the 
negative departure from the WF law is contrary to the theoretical 
predictions for the disordered interacting electron 
systems.\cite{Raimondi, Niven, Catelani} Although one could 
speculate that this unusual breakdown of the WF law in LSCO may 
be related to the strong electron correlations or the 
antiferromagnetic spin fluctuations in cuprates, the validity of 
the WF law in samples with lower Zn-concentrations does not 
support this speculation. Hence, the charge and heat transport 
properties in the moderately disordered, weakly localized regime 
call for deeper theoretical investigations on systems in the 
vicinity of electron localization.

\section{Summary} 

Both the charge and heat transports properties are studied in 
overdoped LSCO where the superconductivity is completely 
suppressed by Zn- or Mg-substitution. When the superconductivity 
is just destroyed with the Zn substitution of $z$ = 0.049 and 
0.082, the ground state is clearly metallic and the WF law is 
found to hold reasonably well, pointing to the Fermi-liquid 
nature of the overdoped cuprates. With higher doping of Zn ($z$ = 
0.13) or Mg ($z$ = 0.15), a low-$T$ resistivity upturn shows up 
but the $T$-dependence of the electrical conductivity follows a 
$T^{1/2}$ law and therefore indicates a metallic ground state. In 
this metallic state, which is in the regime of 3D weak 
localization, the low-$T$ transport data indicate a surprisingly 
negative deviation from the WF law that is opposite to the 
theoretical prediction for the disordered interacting electron 
systems.   

\begin{acknowledgments}

This work was supported by the National Natural Science 
Foundation of China (Grant Nos. 10774137 and 50721061), the 
National Basic Research Program of China (Grant Nos. 2006CB922005 
and 2009CB929502), the Research Fund for the Doctoral Program of 
Higher Education of China (Grant No. 20070358076), and KAKENHI 
Nos. 19674002 and 20030004.

\end{acknowledgments}

\end{document}